# Luminescence and scintillation properties of CsI -- a potential cryogenic scintillator

V. B. Mikhailik[1*], V. Kapustyanyk[2], V. Tsybulskyi[2], V. Rudyk[2], H. Kraus[3]

[1] Diamond Light Source, Didcot, OX11 0DE, UK
[2]Scientific-technical and Educational Centre of low Temperature Studies, I. Franko National University of Lviv, 50 Dragomanova Str., 79005, Lviv, Ukraine
[3]Department of Physics, University of Oxford, Oxford, OX1 3RH, UK

**Abstract**

Caesium iodide is one of the more extensively studied scintillators. Here we present X-ray luminescence spectra, scintillation light output and decay curves as function of temperature, from room temperature down to below 10 K. Features of the observed intrinsic luminescence are explained in terms of radiative recombination of on- and off-centre STE. A model permitting interpretation of the dynamics of luminescence changes in CsI with temperature is suggested. This model includes adiabatic potential energy surfaces (APES) associated with singlet and triplet states of self-trapped excitons (STE) and explains the variation of the luminescence spectra with temperature as a result of re-distribution in the population between on- and off-centre STE. The temperature dependence of the scintillation light yield is discussed in the framework of the Onsager mechanism.

*- corresponding author V.B. Mikhailik, e-mail: vmikhai@hotmail.com, telephone: 01235-778801,

## 1. Introduction

Caesium iodide has been known as a scintillator for over half a century and thus has been studied extensively. When doped with Tl ions it is one of the brightest scintillators, exhibiting a light yield of more than 60,000 ph/MeV at room temperature [1], [2]. A good match of the Tl emission band (maximum at 560 nm) to the spectral sensitivity of photodiodes and a relatively short decay time constant (0.6-0.9 µs) makes this scintillator ideal for detecting ionizing radiation in many applications. These include fundamental [3], [4] and applied research [5], space exploration [6], medical diagnostics [7] and security [8]. Thus, the scintillation properties of this material have been extensively investigated and documented. Pure CsI is also known as a scintillator but exhibits rather different properties. It exhibits fast (~10 ns) emission, peaking at 310 nm, but its scintillation yield at room temperature is very low [9]. However, with cooling of pure CsI the emission intensity increases by more than one order of magnitude. The scintillation light yield at 77 K reaches an impressive 100,000 ph/MeV [10], [11], that makes pure CsI a very attractive scintillator for detector applications at low temperatures.

An opportunity for applying this scintillation material are cryogenic experiments searching for rare events, where a detector, operating in the milli-kelvin temperature range, is used to record simultaneously phonon and scintillation signals [12]. The key features of such phonon-scintillation detectors are low threshold and the ability of event type discrimination that leads to higher sensitivity of the whole experiment [13], [14]. High light yield at cryogenic temperatures is a key requirement for such application and it is quite obvious that pure CsI is an ideal candidate. Recent studies [15] gave promising results for caesium iodide as cryogenic scintillation detector by demonstrating excellent performance of the light channel capable of discriminating two lines (600 eV apart) of an [55]Fe X-ray source. There, the response of the phonon channel was reported to be unexpectedly low. In turn, the resolution of the phonon channel was poor, which limits its usefulness as phonon-scintillation detector. The specifics of the energy transformation at the thermalization stage, controlling the branching between different excitation channels, which eventually dissipates heat and light [16] is the likely cause of this. Therefore it would be important to achieve better insight into these processes in CsI and relate the findings to the performance characteristics of the cryogenic phonon scintillation detector.

There are numerous results on energy structure, luminescence and dynamics of the excitation processes in CsI [17] [18], [16], [19], [20], [21], which provides a wealth of background information for our understanding of the material's scintillation mechanism. The scintillation properties and X-ray luminescence down to 77 K have been documented in a number of papers [10], [11], [22], [23], [24] [25], [26]. However, no data on scintillation properties of CsI below this temperature have been reported hitherto. The lack of systematic investigations of the scintillation characteristics of the crystal below 77 K motivated this study. In this paper we report results on X-ray luminescence, scintillation light output and decay time characteristics of pure CsI over the 6 – 295 K temperature range. The scintillation light yield is the main factor in determining the energy resolution of a light detector. Furthermore, information on the temperature dependencies of luminescence and scintillation decay down to a few Kelvin is essential for insight into the process of transforming absorbed energy into scintillation. The work presented here attempts to generalize these findings, determines the limitations of the performance of cryogenic CsI detectors and assesses possibilities for low-temperature applications.

## 2. Experiment

The samples of pure CsI used in this study were supplied by Hilger Crystals (Margate, UK). For measurements of X-ray luminescence spectra the sample was placed into a closed cycle He cryostat, equipped with a DE-202A cryocooler (Advanced Research Systems). Stabilization of temperature was performed by a Cryocon 32 (Cryogenic Control Systems Inc.) temperature regulator. The emission was excited by a URS-55A X-ray source with Cu-anticathode tube operating at 55 kV and 10 mA. The luminescence spectra were measured using an automated spectrograph M266 equipped with a CCD-camera, incorporating a Hamamatsu S7030-1006S sensor with quartz window sensitive over 200-1100 nm wavelength range. The measured spectra were corrected for the spectral sensitivity of the experimental setup. To study scintillation properties, a sample of dimensions 5×5×1 mm$^3$, polished to optical quality, was placed in a helium constant flow cryostat and excited by α-particles from an $^{241}$Am source. The penetration depth of α-particles in CsI is 25 μm that is deep enough not to be affected by surface quenching observed at the first 10s of nm. Scintillation was detected by a bi-alkali photomultiplier model 9125BQ (Electron Tube Enterprises, Ruislip, UK) sensitive over 200-600 nm wavelength range. For measurements and data analysis we used the multi-photon counting techniques described in detail elsewhere [27].

## 3. Results

### 3.2. Emission properties of CsI over the 10-295 K temperature range

The luminescence studies that are key for understanding the final stage of transforming high-energy excitation into light emission are essential for getting insight into the scintillation process. Therefore we begin with presenting results on luminescence as function of temperature. X-ray luminescence spectra of pure caesium iodide exhibit several emission bands with different dynamics in response to temperature change. Fig. 1 shows luminescence spectra of a CsI crystal under steady-state X-ray excitation as temperature changes. At room temperature emission is relatively weak; only two bands of comparable intensity are observed at 310 and 560 nm. The intensity of the band at 310 nm gradually increases with cooling until about 150 K when a new emission at 340 nm starts to increase. This band totally dominates the emission spectrum of CsI at low temperatures. Fig. 1b also shows two clearly separated peaks at 285 and 295 nm appearing in the luminescence spectra at T< 50 K.

The broad 560 nm band initially shows very little change with cooling but below 150 K the emission intensity rapidly decreases to negligible level. The room temperature luminescence of CsI is known to be very sensitive to impurities: at concentration levels of Tl as low as 1 ppm the intensity of the long wavelength emission in the 400-600 nm range is commensurate with that of the 340 nm band [28]. Such level of impurities cannot be controlled easily and is the usual cause of cross contamination of samples supplied commercially by companies producing both pure and Tl-doped CsI crystals [24], [29]. This emission vanishes with cooling and hence it has no influence on the performance of CsI as cryogenic scintillator. The emission bands observed at low temperature are associated with the intrinsic emission of the crystal.

The nature of intrinsic luminescence of caesium iodide has been studied in detail for several decades and currently there is the consolidated opinion that the main emission features are due to the radiative decay of self-trapped excitons (STE). Two types of STE exist in CsI [30]: the on-centre STE is an electron trapped at a $V_k$-centre, while the off-centre configuration is composed of nearest-neighbour pairs of F- and H- centres. The excited states of two types of STE have singlet and triplet configurations in thermal equilibrium. Thus, the principal features of CsI luminescence can be interpreted in terms of changing populations of these states with time and temperature [17], [31]. The diagram of adiabatic potential energy surfaces (APES) of the STE states displayed in Fig. 2 can be used to illustrate the dynamics of change in CsI luminescence.

The emission band at 340 nm is assigned to the off-centre STE. This band dominates the X-ray luminescence of the CsI spectrum at low temperature and it is well documented in the literature. The band at shorter wavelength (290 nm) is reported as the emission of the on-centre STE. We observed two peaks at 285 and 295 nm in the short-wavelength region of the X-ray luminescence spectra of the CsI crystal used in this study. This observation is an extrinsic feature that is due to the presence of Tl ions in the sample. It is known that thallium exhibits a sharp excitation band at 290 nm in CsI [32] that alters the shape of the measured emission spectrum by creating the trough. The intensity of this band decreases promptly with increase of temperature until it extinguishes at T=50 K. This can be explained as de-population of the excited states of on-centre STE across the small activation barrier that feeds emission of the off-centre STE (see Fig. 2). Manifestation of this effect is a slight increase of the 340 nm emission intensity with increase of temperature from 10 to 50 K (see Fig. 1).

At higher temperatures (>150 K) the process reverses: the thermal activation from the off- to on-centre STE becomes possible and a short-wavelength emission band at 305 nm appears. This band is red-shifted in comparison with the emission of on-centre STE observed at low temperature. Such redshift of the on-centre STE emission band with heating is a common feature of alkali halides [30]. The increase of temperature prompts re-distribution of the initial population of excited states promoting thereby population of on-centre STE levels. Consequently, the intensity of the 305 nm band rises on the expense of the off-centre STE emission band at 340 nm. It should be noted that such re-distribution of excitations between the emitting states is accompanied by the non-radiative thermal quenching that causes a gradual decrease of the integrated emission efficiency with increase of temperature as shown in Fig. 3. The non-radiative decay is due to activation of hopping motion of STE that results in the energy transfer to the quenching centres [21].

### 3.2. Temperature dependence of scintillation properties of CsI

The temperature dependence of scintillation properties of pure CsI is the main interest of this study. We carried out measurements of scintillation properties of pure caesium iodide over the temperature range 6 – 295 K. Despite the low light yield of the crystal at room temperature it was possible to detect a peak due to α-particle excitation in pulse height spectra of CsI. The position of this peak is proportional to the scintillation light output of the crystal. Thus, by measuring the variation of the peak position as a function of temperature one can monitor the change of the light response. The temperature dependence of the light output of the CsI crystal under test is shown in Fig. 3. The scintillation light output exhibits a massive, 15-fold, increase when the crystal is cooled from 295 to 80 K that is consistent with the latest observations of [22]. A trough observed in the light yield versus temperature dependence at around 60 K can be associated with the capture of excited carriers by traps; the measurements of thermoluminescence of pure CsI show glow peaks in this temperature range [29]. When the temperature of the CsI crystal was reduced further down to 20 K we observed an additional increment (ca. 10%) in the light output. Below this temperature the scintillation light yield started to drop, so that at 6 K (the lowest measured temperature) it remains by a factor 14 higher than at room temperature.

The temperature changes of scintillation decay of pure CsI further reflects the complexity of the emission processes in the crystal. It has been the subject of many studies which concentrated on the $10^{-12}$-$10^{-6}$ sec range of time constants [16] [21] [22] [23] [26] [25] [31], [33]. The motivation behind this is fairly obvious as this allows elucidating the thermalization and branching of fractional populations of excited states that determines the features of light emission in the crystal. In turn, investigations of the kinetics of the relaxation process in the microsecond time domain can provide complementary information on the relatively long-lasting dynamics of the energy transfer between different types of excitation centres. The multi-photon counting technique used in this study is best suited for this type of measurement of the decay process in the micro and millisecond range.

Fig. 4 displays changes of scintillation decay curves with temperature. It should be noted that the scintillation decay was measured in the integral regime thus capturing the entire emission spectrum of CsI. Nonetheless, thanks to the data on X-ray luminescence of the crystal under test it is possible to correlate main features of the decay curves with the specific type of emission. At room temperature the decay curves exhibit three markedly different components. Pure CsI is known to feature very fast emission of the order of 10 ns attributed to the emission of on-centre STE [31], although other mechanisms such as recombination of correlated excitations are also considered [26], [33]. Our results show clear evidence of very fast (<0.1 μs) decay that can be attributed to this emission. This component of scintillation event cannot be quantified by our technique due to limited time resolution. We fitted the sum of two exponential decay curves and that gave as best fit values for the decay time constants 1.3 and 28 μs at T=295 K (see fig.5a and Table 1). These are typical values observed for Tl-emission in doped CsI-Tl [1], [24].

When the temperature of the crystal decreases, the first decay time constant remains fairly unaffected while the second increases (see Fig. 6). At T=150 K the long component disappears that can be associated with the extinction of the 500 nm band in the luminescence spectra as can be seen in Fig. 1. Therefore it is sensible to attribute this slow emission component to the luminescence of Tl-centres. Below 150 K the decay kinetics is barely dependent on temperature; the decay curves can be fitted by a single exponential decay with time constant of about 1 μs (Fig.5b, Table 1). That is in good agreement with [31]. Recall, that at this temperature the main intrinsic emission band at 340 nm is due to the singlet transitions in off-centre STE. At 20 K the slow emission component starts to appear and the decay curves must be fitted by a sum of three exponentials (Fig.5c, Table 1). According to [31] the slow decay component is assigned to the transition between excited triplet levels of off-centre STE and the singlet ground state. The triplet levels manifest themselves by slowing the luminescence decay at low temperature when they cannot be de-populated through the thermally activated process [34].

## 4. Discussion

Intrinsic luminescence of pure CsI at cryogenic temperatures features emission of STE. According to Payne [35] the light yield of a scintillator is governed by two competing

processes – exciton-exciton annihilation (Brick mechanism) and electron-hole recombination (Onsager mechanism). The Brick mechanism is the cause of the decrease of light yield through the annihilations of STE which come in contact in the region of high density excitation. In contrast, the Onsager mechanism controls the recombination process of thermalized carriers which are attracted by Coulomb force and form excitons. It has been demonstrated that in alkali halides Brick's mechanism is less efficient compared with other classes of scintillation materials [35] and the Onsager mechanism can facilitate high light yield. The reason for this is that in alkali halides electron mobility is significantly greater than that of a hole due to the effect of hole self-trapping. This results in fast diffusion of electrons and dilution of the excitation density in the active volume where electrons and holes are initially generated by high-energy excitation. Furthermore, in intrinsic scintillators the dominant scattering mechanism is due to electron-phonon interaction. A low optical phonon frequency in CsI (10 meV) means long termalization time and distance. Thus in accordance with MC simulations [36], the thermalization distance of electrons in CsI can be a few hundred nanometres, leading to significant charge separation. Then, the probability of recombination $p$ of thermalized carriers defined by the Onsager mechanism is temperature dependent:

$$p_1 = 1 - \exp(-r_{ONS} / r) \, ,$$

where $r_{ONS}$ is the Onsager radius defined as $r_{ONS} = e^2 / 4\pi\varepsilon_0 \varepsilon k_B T$, with $e$ the electron charge, $\varepsilon$ the static dielectric permeability, $k_B$ the Boltzmann constant and $T$ the temperature. As follows from this equation the efficiency of formation of STE tends to increase with decrease of temperature and distance $r$ between charged particles. This relation defines the temperature dependence of the STE emission in the absence of energy transfer to other emission centres like defects, impurities or activators. Estimates of the value of the Onsager radius in pure CsI at T=10 K gives 130 nm that is consistent with the thermalization distance of electrons in this crystals [36]. Hence, at low temperature a large fraction of thermalized electrons that escaped annihilation have high probability of recombining radiatively with the self-trapped holes, and that results in very high light yield of the crystal.

It is generally accepted that high light yield at low temperature is imperative for good scintillator performance when used as part of a phonon scintillation detector. This is true but to limited extent. A very high scintillation light yield of a material will compromise the performance in phonon channel as the deposited energy is shared between both. It is interesting to examine the balance of energy in CsI. Taking the average energy of an emitted photon as 3.7 eV and adopting 100,000 ph/MeV as the value of light yield at low temperature, one arrives at an impressive 37% energy efficiency for this scintillator. For CaWO$_4$, which is commonly used in cryogenic phonon scintillation detectors this value is closer to 8% [37], 4.6 times less. Thus, even under the assumption of 100% efficiency of phonon detection, the amount of energy that is deposited in the phonon channel of CsI will be reduced and this will adversely affect the energy resolution in the phonon channel.

**Conclusions**

Very high light output of pure CsI at low temperature prompted interest in this material in cryogenic applications. We carried out investigations of X-ray luminescence and scintillation properties of pure caesium iodide from room temperature to 10 K and below. It has been found that the complex nature of the intrinsic luminescence of pure CsI can be explained in the framework of a model of radiative recombination of on- and off-centre STE. The model also allows explaining the features of the luminescence kinetics over a wide temperature range. We also discussed the cause of the exceptionally high scintillation light yield of pure CsI and concluded that this is due to the very large difference in the mobility of electrons and holes that causes prompt spatial separation of hot carriers. This prevents the annihilation of hot excitations by the Brick mechanism. In contrast, the radiative recombination of thermalized electrons and holes governed by the Onsager mechanism is dominant at low temperatures, giving rise to efficient scintillation of CsI. Based on the assessment of energy fractions shared between phonon and light channels we envisage that due to the high scintillation light yield, the amount of energy available for the phonon channel is reduced compared with oxide scintillators, thus having an adverse effect on the energy resolution in the phonon channel.

**Acknowledgment**

The study was supported by a grant from the Royal Society (London) ''Cryogenic scintillating bolometers for priority experiments in particle physics'' and the Science & Technology Facilities Council (STFC).

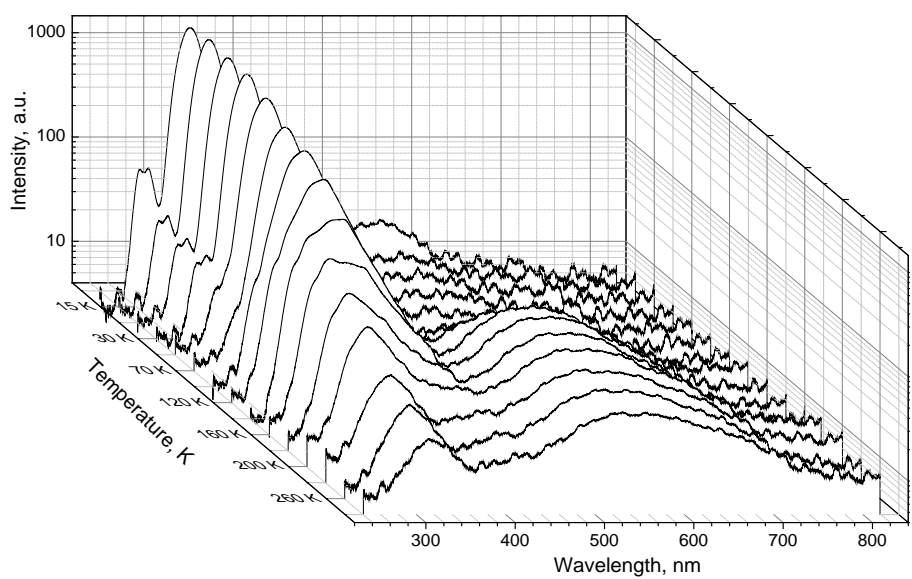

Fig. 1. Isomeric plot showing X-ray luminescence spectra of CsI measured at different temperatures (note that intensity is presented in log scale for the purpose of visualisation).

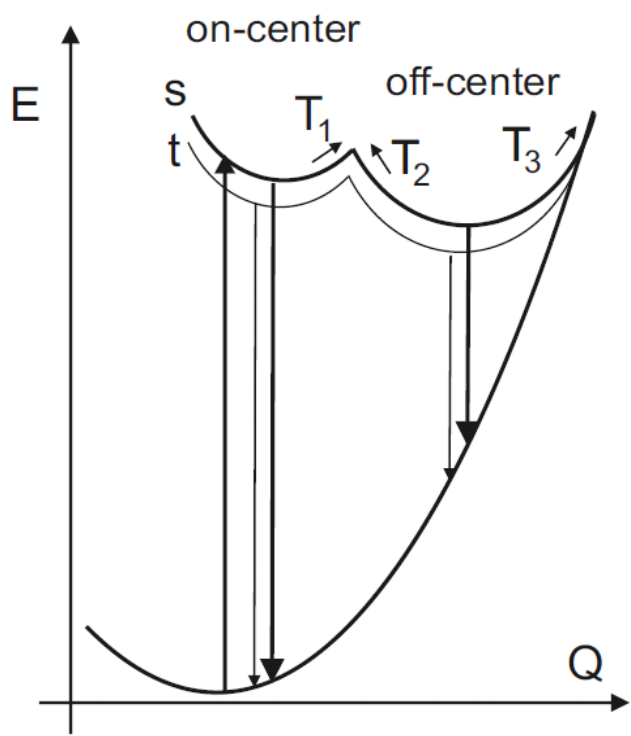

Fig. 2. Scheme of adiabatic potential energy surfaces (APES) of on- and off- centre STE in pure CsI. Vertical arrows represent excitation and emission transitions between singlet (s) and triplet (t) states and the ground state. Small arrows shows the direction of carrier transport occurring between APES at different temperatures ($T_1 < T_2 < T_3$).

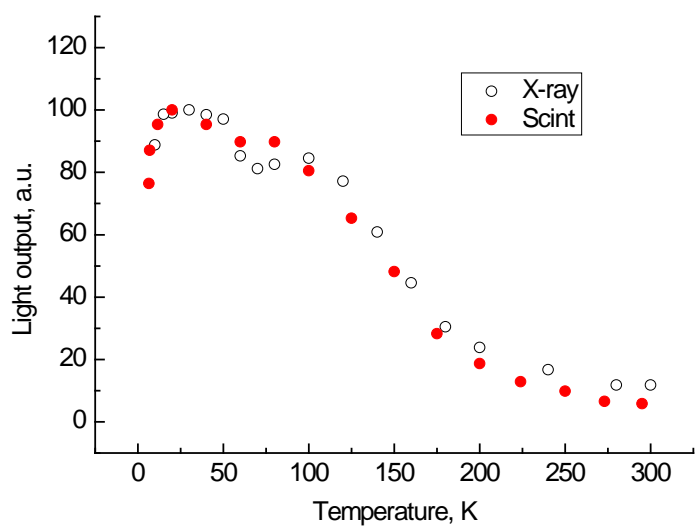

Fig. 3. Light output of CsI as function of temperature for steady-state X-ray and α-particle excitation ($^{241}$Am). The curves are normalised to the maximum value.

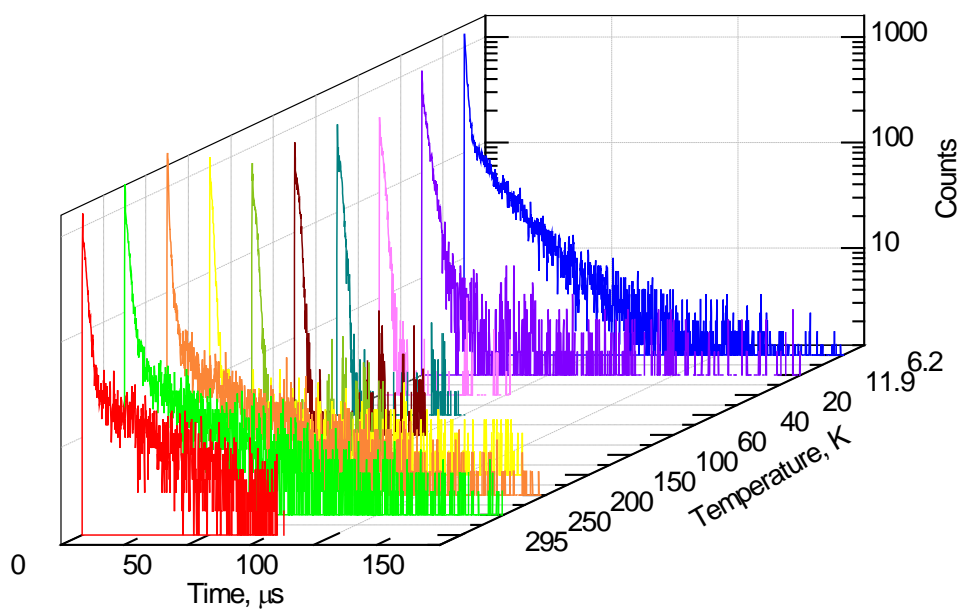

Fig. 4. Isometric plot of decay curves of CsI measured for α-particle excitation ($^{241}$Am ) at different temperatures.

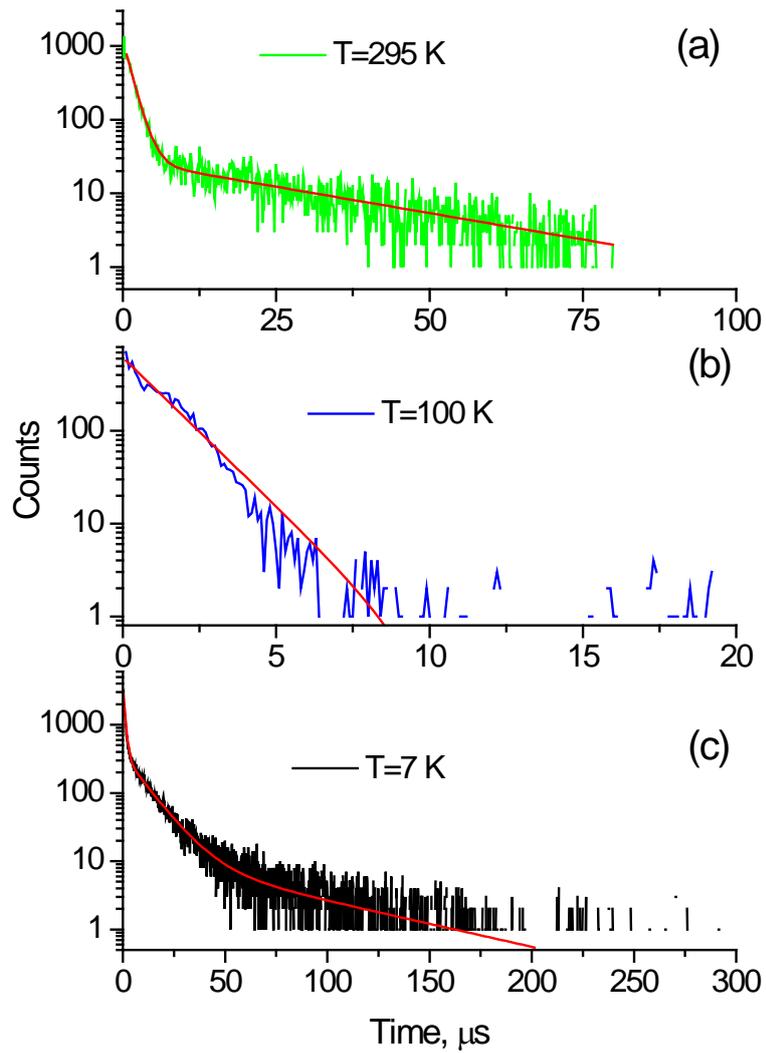

Fig.5 Scintillation decay curves of CsI measured for α-particle excitation ($^{241}$Am ) at different temperatures. Solid lines show shows best fit to the experimental data using two exponential a) T=295 K, one exponential (b) T=100 K and three exponential model (c) T=7 K,  which represents different mechanisms of radiative decay (see text).

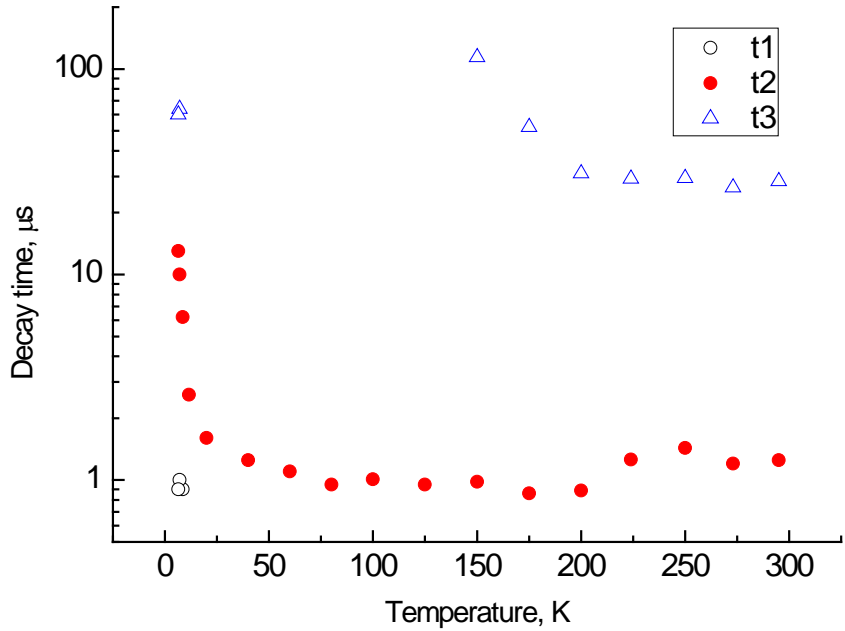

Fig. 6. Temperature dependence of scintillation decay time constants of CsI excited by α-particles from $^{241}$Am.

Table 1

Amplitudes ($A_{1,2,3}$) and decay time constants ($\tau_{1,2,3}$) obtained from multi-exponential fittings of scintillation decay curves of CsI measured at different temperatures and displayed in Fig. 5.

| T, K | $A_1$, % | $\tau_1$, μs | $A_2$, % | $\tau_2$, μs | $A_3$, % | $\tau_3$, μs |
|------|----------|--------------|----------|--------------|----------|--------------|
| 295  | 98       | 1.33±0.01    | 2        | 28.3±1.4     | -        | -            |
| 100  | 100      | 1.36±0.01    | -        | -            | -        | -            |
| 7    | 88.8     | 0.97±0.01    | 10.8     | 10.7±0.2     | 0.04     | 64±10        |